\documentclass[aip,jap,preprint,showpacs]{revtex4-1}
\usepackage{csquotes}
\usepackage{graphicx}
\usepackage{graphics}

\begin{document}
	
	\title{Unconventional domain wall magnetoresistance of patterned Ni/Nb bilayer structures below superconducting transition temperature of Nb}

\author{Ekta Bhatia}
\email[Ekta Bhatia: ]{bhatiaekta@niser.ac.in}
\affiliation{ School of Physical Sciences, National Institute of Science Education and Research (NISER), HBNI, Bhubaneswar, Odisha, 752050, India}
\author{Zainab Hussain}
\affiliation{UGC-DAE Consortium for Scientific Research, Indore Centre, 452017, India}
\author{V. Raghavendra Reddy}
\affiliation{UGC-DAE Consortium for Scientific Research, Indore Centre, 452017, India}
\author{Zoe H. Barber}
\affiliation{Department of Materials Science \& Metallurgy, University of Cambridge, 27 Charles Babbage Road, CB3 0FS, United Kingdom}
\author{Kartik Senapati}
\email[K. Senapati: ]{kartik@niser.ac.in}
\affiliation{ School of Physical Sciences, National Institute of Science Education and Research (NISER), HBNI, Bhubaneswar, Odisha, 752050, India}

\date{\today}

\begin{abstract}

Scattering of spin-up and spin-down electrons while passing through a ferromagnetic domain wall leads to an additional resistance for transport current, usually observed prominently in constricted magnetic structures. In this report we use the resistance of the domain wall as a probe to find an indirect signatures of the theoretically predicted spin-singlet supercurrent to spin-triplet supercurrent conversion effect of ferromagnetic domain walls.  Here we examine the domain wall induced resistance in Ni stripe in a bilayer Ni/Nb geometry in the normal state and in the superconducting state of Nb. By making a 6$\mu$m wide gap in the top Nb layer we routed the transport current through the Ni layer in the normal state and in the supercondcuting state of Nb. In the normal state of Nb, in-field transport measurements showed a clear domain wall magneto-resistance (DWMR) peak near the coercive field, where the domain wall density is expected to be maximum. Interestingly, however, below the superconducting transition temperature of Nb, the DWMR peak of the Ni layer showed a sharp drop in the field range where the number of domain walls become maximum. This observation may be a possible signature of magnetic domain wall induced spin-triplet correlations in the Ni layer due to the direct injection of spin-singlet Cooper pairs from Nb into the magnetic domain walls. 

\end{abstract}

\maketitle
\section{Introduction}
 Spin singlet Cooper pairs, when injected across a superconductor(S)-ferromagnet(F) interface, decay within a distance of a few nm in the ferromagnet due to the strong magnetic exchange experienced by the Cooper pairs\cite{2}. Therefore, a complete synergy between superconductor and spintronics\cite{3,4} is possible only through the generation of long-range, spin triplet Cooper pairs at carefully engineered S-F interfaces. It has been predicted more than a decade ago that magnetic inhomogeneity at the S-F interface is the key ingredient for generation of triplet supercurrent\cite{7} from singlet supercurrent. In S-F systems with homogeneous ferromagnetic interface, Cooper pairs undergo spin mixing \cite{3,4} giving rise to short range S$_{z}$ = 0 singlet and triplet components with a proximity length of the order of 1 to 10 nm. Introduction of magnetization non-collinearity at the S-F interface leads to spin rotation\cite{3,4} along with spin mixing which converts short range S$_{z}$ = 0 triplet component into the long range S$_{z}$ = $\pm$1 triplet component. When a Cooper pair with S$_{z}$ = $\pm$1 triplet component propagates through a ferromagnetic layer, the exchange field of ferromagnet no longer has a pair-breaking effect on it\cite{3,4}. 
 
In this direction, a long range supercurrent was reported in Josephson junctions with half metallic ferromagnetic CrO$_{2}$ barriers\cite{10}. However, the results were less reproducible due to the random nature of inhomogeneous magnetic states at the SF interface. Later on, a series of experiments were reported demonstrating the generation of triplet supercurrent in SFS Josephson junctions by carefully engineering the SF interface to be magnetically inhomogeneous\cite{11,12,14,17}. Recently, a series of experiments have been carried out to generate triplet supercurrent in Josephson junctions with textured ferromagnets\cite{22}, bilayer and trilayer ferromagnetic regions\cite{24}, spin injection\cite{25}, and via spin-active
interfaces\cite{26}, by creating a misalignment of magnetic moments at the SF interface. Evidences of induced triplet superconductivity have also been reported using scanning tunneling spectroscopy of SF bilayers\cite{27}, Andreev spectroscopy in SF junctions\cite{28}, conductance measurements\cite{28},  and critical temperature measurements of S-F spin-valve\cite{33}. In a recent report by Robinson et al.\cite{36}, an in-plane Bloch domain wall was created artificially in Gd by using non-parallel alignment of the Ni layer moments in a Ni-Gd-Ni trilayer in the Nb-Ni-Gd-Ni-Nb Josephson junction. An enhancement of supercurrent was observed in this structure by modifying the low temperature DW state in Gd. The magnetic inhomogeneity has been created artificially in all these reports. However, theoretically, a magnetic domain wall (DW) at the SF interface should also be able to produce triplet Cooper-pairs\cite{34,35,7,4}. The fundamental difficulty in implementing this concept is that a domain wall exists only at the interface of two ferro-magnetic domains. In order to utilize the domain wall as a singlet to triplet converter, singlet Cooper pairs need to be passed from one side of the domain wall to the other side, whereas the ferromagnetic exchange field in the domains forbids the presence of spin-singlet Cooper pairs. 

In this report, we have tried to inject spin-singlet Cooper pairs directly into the domain walls of a ferromagnetic layer from a superconducting layer deposited on top of the ferromagnetic layer in a bilayer stripe geometry as shown in Fig. 1(a). It is well known \cite{38} that domain walls in a ferromagnet add a resistance to the overall resistance of the ferromagnetic layer. This domain wall magnetoresistance (DWMR) becomes maximum in the magnetic field range where the number of domain walls maximizes. We find that in the normal state of the Nb layer the usual domain wall magnetoresistance appears superposed on the anisotropic magnetoresistance of Ni layer. However, in the superconducting state of Nb we observe an unconventional decrease in DWMR in the field range where the number of domain walls become maximum. This observation may be a signature of the singlet-triplet conversion effect of the domain walls. 

\section{Experimental Details}
Nb-Ni thin films with thickness of 55 nm and 100 nm were prepared at room temperature using dc-magnetron sputtering of high purity($99.999\%$) niobium and nickel targets on cleaned Si-SiO$_{2}$ substrates. Deposition of the films was carried out at a base pressure of $1 \times 10^{-9}$ mBar in an ultra high vacuum chamber. Optical lithography and reactive ion etching techniques were used to fabricate the Nb-Ni-Nb planar structures as shown in Fig 1 (a) . Four probe electrical transport measurements were carried out using a liquid cryogen free dewar. The voltage was measured by reversing the current and averaging over five measurements. The temperature was stable within a range of $\pm$3 mK.  Magnetization measurements were carried out using vibrating sample magnetometer. The magnetic domains in patterned structures were imaged
by magneto-optical Kerr microscope in longitudinal mode for an in-plane magnetic field applied along the length and width of the stripes.
\section{Results}
\subsection{Domain wall magneto-resistance in Nb/Ni stripes in the normal state of Nb:}
\begin{figure}[htbp!]
	\includegraphics[width=12cm]{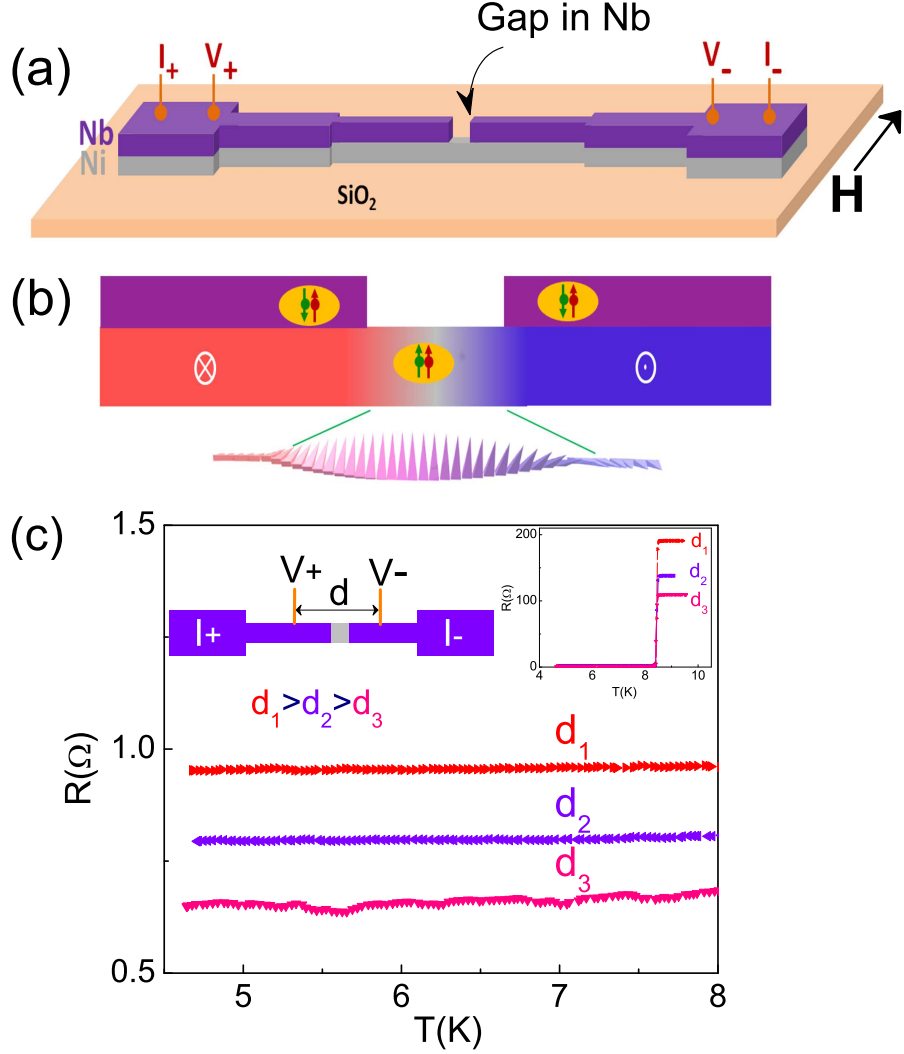} 
	\caption{\textbf{Schematic measurement geometry and singlet-triplet conversion by domain walls} (a) A gap of dimensions 3 $\mu$m was created in a Ni/Nb bilayer stripe of width 6 $\mu$m. (b)The zoomed view of Bloch domain wall shows the non-collinear magnetic structure through which the singlet Cooper-pairs pass when injected into the domain walls.	(c) The resistance in the superconducting state of Nb contacts have been measured in a similar stripe for various separations of voltage contacts (d) keeping the current contacts fixed. We observe that for larger separation of voltage contacts, the residual resistance in the superconducting state of Nb is larger indicating that the diffusion of current from superconducting layer to ferromagnetic layer happens beyond the gap regime also.}
\end{figure}  
We have lithographically patterned stripes of width 6 $\mu$m from Ni/Nb bilayer thin films as shown in the schematic of Fig.1(a). Ni has been chosen as the ferromagnet layer because of its high exchange field and the consequent low singlet pair coherence length ($\sim$4 nm)\cite{54}. The thickness of nickel layer was chosen as 100 nm, which, from earlier reports\cite{37,55}, exhibits Bloch domain walls. A 3 $\mu$m wide gap in the top Nb layer, as shown in the Fig.1 (a), was created in the central part of the stripe which was 6 $\mu$m wide. The large gap in the Nb layer excludes any possibility of Josephson coupling between the two Nb ends through singlet supercurrent. Fig. 1(b) shows the schematic diagram of a Bloch domain wall in the Ni layer, emphasizing the magnetic inhomogeneity. As discussed earlier, the gap in the superconducting Nb layer promotes diffusion of Cooper pairs into the ferromagnetic Ni layer at temperatures below superconducting transition (T$_c$), as shown in Fig. 1(b). In the absence of the gap, the superconductor would completely bypass the Ni layer, as shown in the supplementary Fig. S1. In fact, we find that the geometrical gap in the superconducting layer promotes diffusion of Cooper pairs well beyond the gap dimensions. In Fig. 1(c), we show the residual resistance in the superconducting state of such a Nb/Ni stripe with voltage measured at three different separations (varying between 1mm and 1.5 mm) in the gap region, keeping the bias current contacts unchanged as shown in the inset of Fig. 1(c). Transition temperature in all cases was 8.5 K as shown in another inset of Fig. 1(c). Clearly, larger separation of the voltage contacts leads to a larger resistance below the transition temperature of Nb. One can argue that the observed effect is due to charge imbalance, which is a nonequilibrium phenomenon arising when the quasiparticle current in a normal metal converts to supercurrent in a superconductor. According to previous reports\cite{58,59}, the characteristic feature of charge imbalance is an increase in resistance below the superconducting transition which is not the case here. Therefore, the possibility of charge imbalance is excluded in the present experimental geometry. Although the major part of the resistance below T$_c$ of Nb, in this geometry, comes from the Ni in the gap region, Fig. 1(c) shows that there is some contribution to resistance from Ni beyond the gap indicating diffusion of Cooper pairs beyond the gap region.  
\begin{figure}[t]
	\includegraphics[width=12cm]{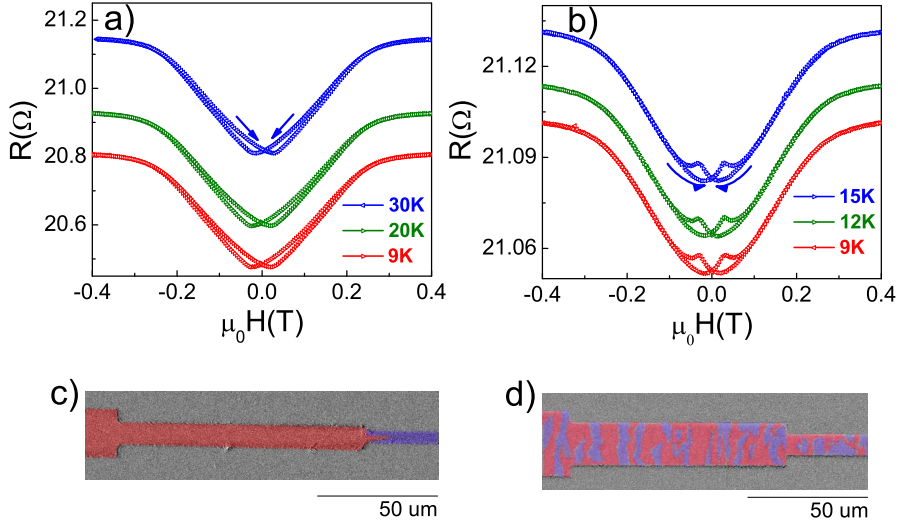}
	\caption{\textbf{Domain wall magneto-resistance of Ni/Nb planar structures for temperatures above T$_{c}$ of Nb.}
		(a) MR curves for magnetic field applied along the length of stripes show the intrinsic magnetoresistance of nickel for different temperatures above T$_{c}$.
		(b) MR curves for magnetic field applied perpendicular to the length of stripes show the domain wall magnetoresistance (DWMR) in addition to intrinsic MR of nickel. The MR curves are similar for different temperatures above superconducting transition for magnetic field applied along and perpendicular to the length of stripes.
		(c) Longitudnal Kerr microscope image for magnetic field applied parallel to the length of stripes of nickel shows the two domain walls propagating towards the centre. 
		(d) Kerr microscope image for magnetic field applied perpendicular to the length of stripes shows the multiple domain walls in nickel. Both the Kerr measurements have been done at room temperature. Red color and blue color corresponds to the domains orienting in opposite direction.}
\end{figure}
In Fig. 2, we present the magneto-transport behavior of the Ni/Nb planar structures at temperatures above the superconducting transition in presence of an in-plane applied magnetic field. These measurements are performed for a sample different from the one shown in Fig. 1(c). Panel 2(a) shows the response for magnetic field along the length while panel 2(b) shows the response in a field along the width of the stripe. Henceforth, keeping in mind that magnetic field is always applied in the plane of the substrate, throughout the text, we will refer to these two field configurations as longitudinal and transverse field, respectively. Intrinsic magneto-resistance of nickel was observed in longitudinal field (In Fig 2(a)), where the MR minima typically corresponds to the  coercive field\cite{52}. This was found to be same irrespective of change in temperature value up to 30 K, due to very high Curie temperature of nickel. In transverse field configuration, MR peaks were observed with peak amplitude of 5.9 $m\Omega$ at the coercive field, as shown in Fig 2(b). The peak amplitude was found to be same for various temperatures up to 15 K. MR hysteresis was observed in both the configurations which is, however, an expected effect due to the magnetic hysteresis of the Ni layer. The difference is the additional peak in the MR hysteresis in the transverse field configuration when we have an order of magnitude more number of domain walls. The observed peak cannot be from anisotropic magnetoresistance (AMR) of Ni film in this case because (i) we have not observed the additional peak in the MR hysteresis of Ni thin films or in Ni/Nb bilayers in longitudinal or transverse field configurations, as shown in the supplementary Fig. S2 and (ii) the amplitude of DWMR peaks shown in Fig. 2(b) are temperature independent, which is unlike the behavior of AMR which is usually a temperature dependent quantity. We note that the peaks in MR, in transverse configuration, appear at field values corresponding to the minima in MR in longitudinal configuration, shown in panel 2(a). Usually, in a ferromagnetic film, the number of domain walls is maximum near the coercive field. Therefore, the origin of MR peaks in transverse configuration can be correlated to the number of domain walls. In order to verify this, we performed Kerr microscopy of nickel stripes of similar thickness in longitudinal and transverse configurations. As shown in panel 2(c), only two domain walls were observed in longitudinal configuration for fields applied near the coercive field. On the other hand, large number of domain walls were observed in transverse configuration near the coercive field as shown in panel 2(d). Therefore, in the transverse configuration, the DWMR contributions of a large number of domain walls add up to give the observed peaks in the MR curves. Whereas in the longitudinal configuration, due to the low number density of domain walls, no DWMR peaks were observed. The observation of a clear DWMR peak in the transverse field configuration ensures that, in this measurement geometry, an injected current passes through a large number of domain walls in the nickel stripe. Therefore, below the T$_c$ of Nb in the gapped Ni/Nb bilayer stripes, it is very likely that a significant number of diffused singlet Cooper pairs will encounter a domain wall directly. Therefore, this is a suitable measurement configuration to look for any signatures of singlet to triplet conversion through domain walls upon decreasing the temperature below superconducting transition.



\subsection{Magneto-transport below superconducting transition temperature on Nb}
In Fig. 3, we present the magneto-transport behavior of the Ni/Nb planar structures at temperatures below the superconducting transition in the transverse configuration. Fig. 3(a) shows the magnetoresistance (MR) curve measured at 8 K which is very close to superconducting transition (T$_c$ $\sim$8.3 K). Two distinct MR peaks were observed whose amplitudes were found to be more than 50 times higher than the normal state DWMR shown in Fig. 2(b). Since this measurement was done very close to the superconducting transition, the Nb layer is expected to respond to domain wall stray fields very sensitively. Therefore, the origin of MR peaks observed at this temperature is the usual suppression of superconductivity due to the out of plane component of domain walls present in nickel as observed in earlier reports\cite{44, Ekta}. Since the overall domain wall stray field maximizes near the coercive field, the MR peaks, in this case appear near the coercive field. At lower temperatures, superconductivity is less sensitive to the out of plane stray field of nickel domain walls. Therefore, these peaks indicating suppression of superconductivity should diminish at lower temperatures. This is what we observe in the MR data measured at 7 K, as shown in Fig. 3(b). Moreover, we observe a DWMR peak with peak amplitude of 0.2 $m\Omega$ at 7 K as shown in the inset of Fig. 3(b).

\begin{figure}[b]
	\includegraphics[width=12cm]{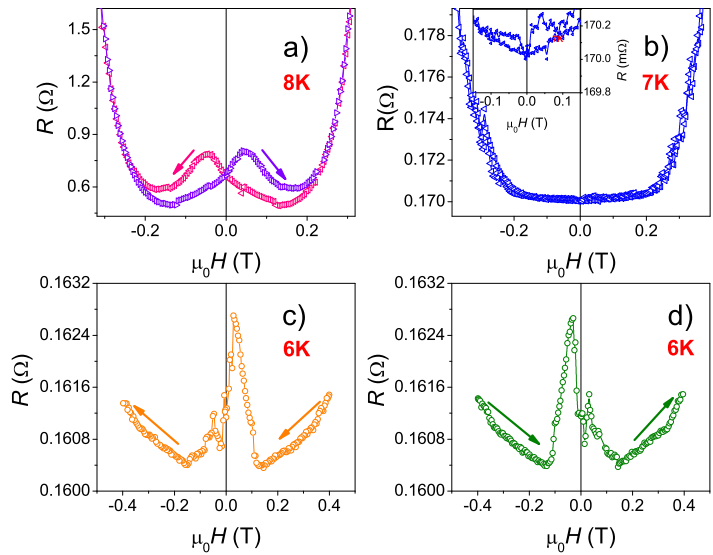}
	\caption{\textbf{Comparison of Magneto-transport(MR) measurements below the superconducting transition in transverse field configuration.}
		(a) The MR measurements at 8K follows the magnetization curve shown in supplementary Fig.1(c). The variation of stray field of domain walls of nickel is reflected in the MR curve.
		(b) The MR measurements at 7K shows the reduction of stray field effect of domain walls on superconductivity of niobium.
		(c) The MR measurements at 6K shows a decrease in resistance at positive field on sweeping the field from positive to negative.
		(d) The MR measurements at 6K shows a decrease in resistance at negative field on sweeping the field from negative to positive.}
\end{figure}

Fig. 3(c) and Fig. 3(d) show the MR measurements at a temperature of 6K with magnetic field swept in negative (+ve to -ve field) and positive (-ve to +ve field) directions, respectively. Since the effect of domain wall stray field was minimized below 7 K as shown in Fig.3(b), the DWMR effect was possible to observe. It can be seen in Fig. 3(c) that while decreasing the magnetic field from positive saturation the DWMR effect sets in as soon as domain formation starts, leading to increase in resistance. However, as the number of domain walls become significant, the resistance starts to decrease. Fig. 3(d) shows the same feature while ramping the field from negative saturation towards positive saturation. The spikes represent jumps in resistance due to abrupt re-configuration of vortices as reported earlier\cite{44}. The observed spikes cannot originate from the movement of the magnetic domain walls that reside in the FM layer under the action of the applied current. The reported values\cite{62} of current density which leads to domain wall motion is  of the order of 10$^{8}$ A/cm$^{2}$, which is many orders of magnitude higher compared to the current densities used in this case ($\sim$10$^{4}$ A/cm$^{2}$). Therefore, the possibility of magnetic domain wall motion is excluded in the present work. We observe that the DWMR at 6 K is more compared to that at 7 K. Earlier report on Nb/Garnet bilayer has shown that superconductivity shrinks the domains\cite{49}. Therefore, the increase in DWMR below 7 K may be due to an increased number of domain walls, as reported earlier\cite{49}. However, the unconventional drop in resistance can not be explained due to change in domain structure with temperature. The field dependent decrease in resistance can not be explained in terms of Nb-Ni proximity effect, because, it should be present at all field values. By studying the conductance-voltage characteristics, one can study the proximity effect at the interface of S-F hybrids\cite{60,61}, which is not the purpose of the present work. Rather, the purpose of this work is to look at the changes in DWMR due to a direct proximity of the domain walls with the Cooper pairs, which is possible to measure via magneto-transport measurements. Here we note that unlike planar Josephson junctions, the large ferromagnetic gap of $\sim$3 $\mu$m, in this case, forbids any direct coupling between the superconducting electrodes at the gap.    

\section{Discussion}

In order to understand the suppressed DWMR effect in certain field range, in Fig. 4, we schematically show the different resistances seen by the current at temperatures above and below the superconducting transition. For temperatures above the superconducting transition, both Nb and Ni layers are normal metals, and they carry the total current in parallel according to their resistance values. A current going through the Ni layer encounters the domains and domain walls of nickel. Therefore, the effective resistance seen by the current in Ni layer consists of resistance of domains $R_{D}$ and domain walls $R_{DW}$.

\begin{figure}[htbp!]
	\includegraphics[width=12cm]{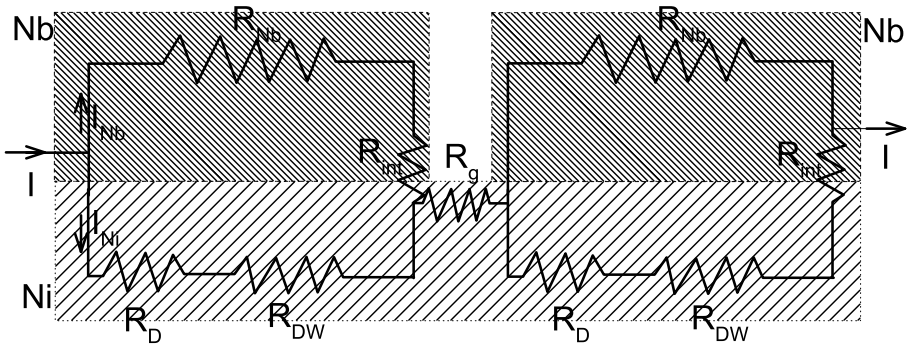}
	\caption{\textbf{Schematic parallel resistance model.}
Above the critical temperature of Nb, The total current I flows in parallel through the Nb layer(I$_{Nb}$) and in the Ni layer(I$_{Ni}$). In the superconducting state of Nb, R${Nb}$ becomes zero and the measured resistance gets contributions from (i) Domain resistance R$_D$, (ii) Domain wall resistance R$_{DW}$, (iii) Nb-Ni Interface resistance R$_{Int}$, and (iv) Resistance of the Ni present in the gap R$_g$.}
\end{figure}
At temperatures below the superconducting transition the resistance, $R_{Nb}$, of the niobium electrodes become zero. The singlet Cooper pairs, which diffuse directly into the domain walls may convert into triplet pairs\cite{7,34,35}.  The spin-diffusion length of Ni\cite{56} ($\sim$21$\pm$2 nm) limits the decay length of triplet Cooper pairs beyond 23 nm, as predicted theoretically\cite{57}. Due to the long-range nature, the triplet Cooper pairs may reduce the effective resistance observed in Fig. 3(c) and 3(d) by reducing the domain resistance R$_D$. The comparison of supplementary Fig. S3(a) and Fig. S3(b) shows that number of domain walls increases as one decreases the field from saturation and approaches the coercive field. We believe that increased number of domain walls increases the chances of direct Copper pair injection into the domain walls leading to an increase in singlet-triplet conversion and hence a decrease in resistance as the number of domains become significant. We must, however, discuss a few other possible scenarios which may lead to such anomalous decrease in magnetoresistance. In an earlier report \cite{46} on Nb/BaFe$_{12}$O$_{19}$ bilayers the authors have ascribed sharp dips in MR to the onset of domain and domain wall superconductivity. However, we must point out that these measurements were done close to the transition temperature where domain and domain wall stray field can have a large effect. Therefore, this possibility can be excluded in our case because, our data (Fig.3) shows that below 7 K the stray field of domain walls no longer has any effect on the MR. Another possible effect that may cause an apparent decrease in magneto-resistance in hybrid S-F systems is the vortex locking effects \cite{53}. In order to rule out this possibility we have measured the DWMR in the superconducting state with varying bias current. With increasing bias current, the Lorentz force on any possible vortices in the Nb layer would increase. Therefore, if vortex locking would have been the cause of the observed resistance drop in the DWMR peaks (Fig.3(c)), the corresponding field values would have been different at all bias currents. However, we find that the decrease in DWMR starts around the same field value for all bias currents (shown in the Supplementary Fig. S4). This observation indicates that the decrease in DWMR is connected to the magnetization dynamics of the Ni layer which remains unaffected by the bias current.

\section{Conclusion}
In conclusion, we have studied DWMR for temperatures above and below the superconducting transition in a Ni/Nb planar structure. In the Ni/Nb bilayer stripe geometry, by carving a gap in the Nb layer, we are able to measure the DWMR of the Ni layer below T$_c$. In the conventional S-F bilayer geometries, used in the earlier reports, current is predominantly confined to the superconducting layers and the intrinsic DWMR property would not be possible to measure. We observed unconventional drop in DWMR of Ni layer at temperatures below the superconducting transition of Nb layer, which was not present in the normal state of the Nb layer. We believe that our measurement geometry allowed for direct injection of singlet Cooper pairs into the domain walls which resulted in singlet-triplet conversion leading to a decrease in the domain resistance. Since this process is dependent on the number of domain walls, the effect was apparent in the magnetic field range where number of domain walls maximized. We have excluded other possibilities leading to the observation of similar drop in MR, such as domain and domain wall superconductivity and vortex locking-unlocking effects. Although the triplet super-current generated by domain walls is not directly accessible, our experiment provides a possible indirect signature of the theoretically predicted singlet-triplet conversion effect of domain walls. This experiment may motivate measurement of planar Josephson junctions with a magnetic gap which accommodates a single pinned domain wall which would be useful for the field of superconducting spintronics. 

\section{Acknowledgements}
KS acknowledge the funding from National Institute of Science Education and Research(NISER), DST-Nanomission (SR/NM/NS-1183/2013) and DST SERB (EMR/2016/005518) of Govt. of India. 


\begin{thebibliography}{1}\footnotesize

 \bibitem{2} A. I. Buzdin, Rev. Mod. Phys. \textbf{77}, 935 (2005). 
\bibitem{3} J. Linder, and J. W. A. Robinson, Nat. Phys. \textbf{11}, 307–315 (2015).
\bibitem{4} M. Eschrig, Rep. Prog. Phys. \textbf{78}, 104501 (2015). 

\bibitem{7} F. S. Bergeret, A. F. Volkov, and K. B. Efetov, Phys. Rev. Lett. \textbf{86}, 4096 (2001).

\bibitem{10} R. S. Keizer, S. T. B. Goennenwein,  T. M. Klapwijk, G. X. Miao, G. Xiao, and A. Gupta, Nature \textbf{439}, 825–827 (2006).
\bibitem{11}  I. Sosnin, H. Cho, V. T. Petrashov, and A. F. Volkov, Phys. Rev. Lett. \textbf{96}, 157002 (2006).
\bibitem{12} J. W. A. Robinson, J. D. S. Witt, and M. G. Blamire, Science \textbf{329}, 59–61 (2010).
\bibitem{14}  T. S. Khaire,  M. A. Khasawneh, W. P. Pratt, and  N. O. Birge, Phys. Rev. Lett. \textbf{104}, 137002 (2010).
\bibitem{17} C. Klose, T. S. Khaire, Y. Wang, W. P. Pratt, Jr., N. O. Birge, B. J. McMorran, T. P. Ginley, J. A. Borchers, B. J. Kirby, B. B. Maranville, and J. Unguris, Phys. Rev. Lett. \textbf{108}, 127002 (2012).
\bibitem{22} I. Eremin, F. S. Nogueira, and R-J. Tarento, Phys. Rev. B \textbf{73}, 054507 (2006).
\bibitem{24} L. Trifunovic, Z. Popovic, and Z. Radovic, Phys. Rev. B \textbf{84}, 064511 (2011).
\bibitem{25} A. G. Mal\'{s}hukov, and  A. Brataas, Phys. Rev. B \textbf{86}, 094517 (2012).
\bibitem{26} M. Eschrig, and T. L\"{o}fwander, Nat. Phys. \textbf{4}, 138143 (2008).
\bibitem{27}  Y. Kalcheim, T. Kirzhner, G. Koren,and O. Millo, Phys. Rev. B \textbf{83}, 064510 (2011).
\bibitem{28}  C. Visani, et al. Nat. Phys. \textbf{8}, 539 (2012).
\bibitem{33} P. V. Leksin, N. N. Garif’yanov, I. A. Garifullin, Y. V. Fominov, J. Schumann, Y. Krupskaya, V. Kataev, O. G. Schmidt, and B. B\"{u}chner, Phys. Rev. Lett. \textbf{109}, 057005 (2012).
\bibitem{36} J. W. A. Robinson, F. Chiodi, G. B. Hal\'{a}sz, M. Egilmez, and M. G. Blamire, Sci. Rep. \textbf{2}, 1-6 (2012).
\bibitem{34} Y. V. Fominov, A. F. Volkov, and K. B. Efetov, Phys. Rev. B \textbf{75}, 104509 (2007).
\bibitem{35} D. Fritsch, and J. F. Annett, Supercond. Sci. Technol. \textbf{28}, 085015 (2015).
\bibitem{38} J. Ieda, S. Takahashi, M. Ichimura, H. Imamura and S. Maekawa, J. Magn. Magn. Mater. \textbf{310}, 2058 (2007).
\bibitem{54} J. W. A. Robinson, S. Piano, G. Burnell, C. Bell, M. G. Blamire, Phys. Rev. B \textbf{76}, 094522 (2007).
\bibitem{37}  C. T. Hsieh, J. Q. Liu, and J. T. Lue, Appl. Surf. Sci. \textbf{252}, 1899 (2005).
\bibitem{55} J. Silcox, Phil. Mag. \textbf{8}, 7 (1963). 
\bibitem{58} P. C.-Zimansky, Z. Jiang, and V. Chandrasekhar, Charge imbalance, New J. Phys. \textbf{9}, 116 (2007).
\bibitem{59} K. Y. Arutyunov, S. A. Chernyaev, T. Karabassov, D. S. Lvov, V. S. Stolyarov and A. S. Vasenko, J. Phys.: Condens. Matter \textbf{30}, 343001 (2018).
\bibitem{52} R. T. McGuire, and R. I. Potter, IEEE Trans. Magn. \textbf{11}, 1018 (1975).
\bibitem{44} E. J. Pati\~{n}o, C. Bell, and M. G. Blamire, Eur. Phys. J. B \textbf{68}, 73-77 (2009).
\bibitem{Ekta} E. Bhatia,  Z. H. Barber, I. J. Maasilta, and K. Senapati, AIP Adv. 9, 045107 (2019).
\bibitem{62} M. Kl\"{a}ui, C. A. F. Vaz, J. A. C. Bland, W. Wemsdorfer, G. Faini, E. Cambril, L. J. Heyderman, F. Nolting and U. R\"{u}diger, Phys. Rev. Lett. \textbf{94} 106601 (2005). 
\bibitem{49}  V. V. Vlasov, A. I. Buzdin, A. Melnikov, U. Welp, D. Rosenmann, L. Uspenskay, V. Fratello, and W. Kwok, Phys. Rev. B \textbf{85}, 064505 (2012).
\bibitem{60} P. Charlat, H. Courtois, Ph. Gandit, D. Mailly, A. F. Volkov, and B. Pannetier, Phys. Rev. Lett. \textbf{77}, 4950 (1996).
\bibitem{61} M. Giroud, H. Courtois, K. Hasselbach, D. Mailly, and B. Pannetier, Phys. Rev. B \textbf{58}, R11872 (1998).
\bibitem{56} C. E. Moreau, I. C. Moraru, N. O. Birge, and W. P. Pratt, Appl. Phys. Lett. \textbf{90}, 012101 (2007).
\bibitem{57} C. Cirillo, S. Voltan, E. A. Ilyina, J. M. Hernandez, A. Garcia-Santiago, J. Aarts, and C. Attanasio, New J. Phys. 19, 023037 (2017).
\bibitem{46}  Z. Yang, M. Lange, A. Volodin, R. Szymczak, and V. V. Moshchalkov, Nat. Mater. \textbf{3}, 793 (2004). 
\bibitem{53} A. Yu. Aladyshkin, A. V. Silhanek, W. Gillijns and V. V. Moshchalkov, Supercond. Sci. Technol, \textbf{22}, 053001 (2009).





\end{thebibliography}

	\clearpage
   


\end{document}


\maketitle
\begin{center}
{\bf Supplementary information\\ Unconventional domain wall magnetoresistance of patterned Nb/Ni bilayer structures below superconducting transition temperature of Nb.\\ Bhatia et al. \\}
\end{center}

\section{Magnetoresistance of Ni-Nb bilayer without any gap in Nb layer}
In the supplementary Fig. 1, we present the magneto-transport behavior of the Nb/Ni patterned bilayer (without any gap in the Nb layer) at temperatures above and below the superconducting transition (T$_{c}$) in presence of an in-plane applied magnetic field. We observed DWMR peaks at coercive fields for temperature above T$_{c}$ as shown in Fig. 1(a). In the superconducting state of Nb layer, no changes in resistance were observed as a function of an in-plain applied field as shown in Fig. 1(b). This shows that the superconductor completely bypasses the Ni layer and no features of DWMR are present below T$_{c}$. Therefore, a gap in Nb layer is needed to study DWMR for temperatures below T$_{c}$.

\begin{figure}[h]
	\includegraphics[width=12 cm]{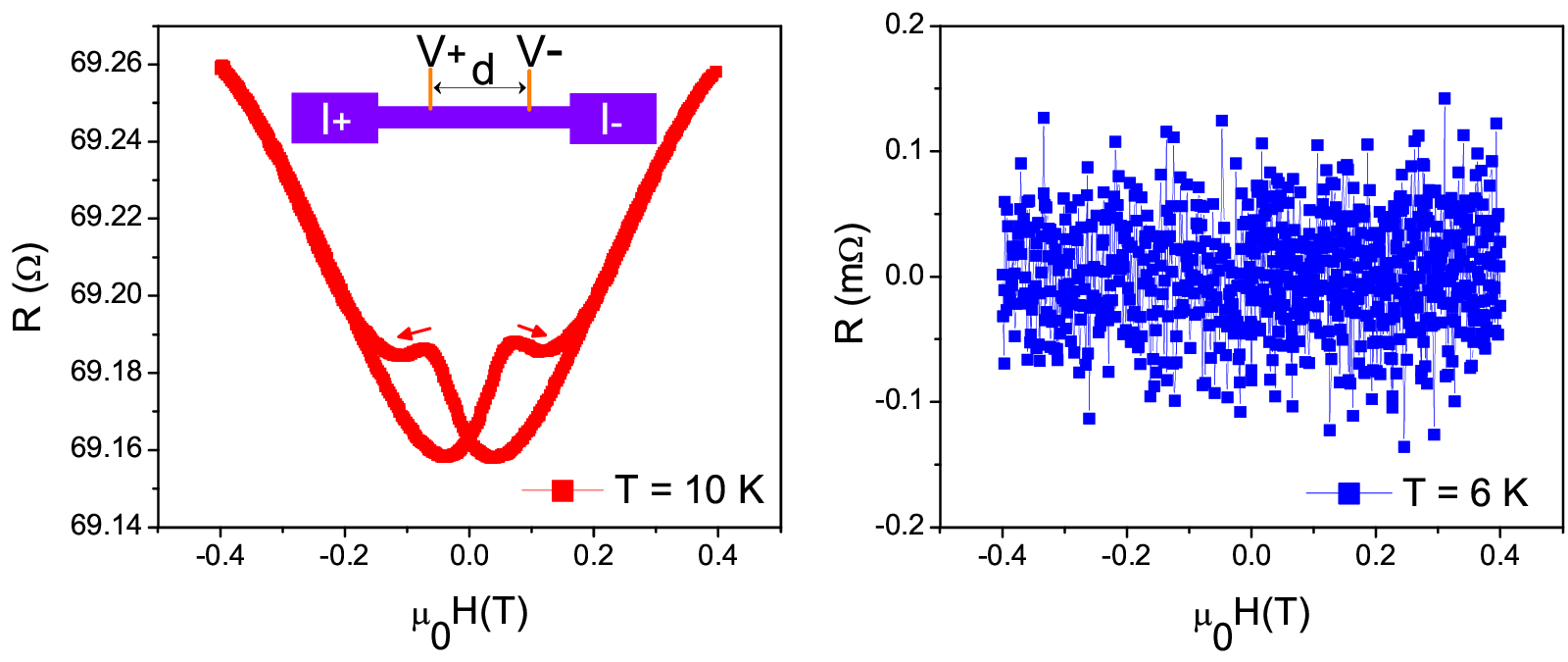}
	\caption{\textbf{ Magneto-transport(MR) measurements of control sample (Nb/Ni patterned bilayer without having any gap) for temperatures above and below the superconducting transition.}
		(a) The MR measurements at 10K show the domain wall magnetoresistance (DWMR) in addition to intrinsic MR of nickel, Inset shows the meassurement geometry for the Nb/Ni patterned bilayer.
		(c) The MR measurements at 6K shows zero resistance for all field values ranging from 0.4 T to -0.4 T.	}
\end{figure}

\section{Anisotropic magnetoresistance of Ni and Ni/Nb bilayer films}

In order to show that the additional peak appearing in the magnetoresistance curve in the transverse field orientation is indeed due to domain wall magnetoresistance (DWMR), we have measured the longitudinal (current along the applied magnetic field direction) and transverse (current at 90 degrees with respect to the magnetic field) magnetoresistance of a 100nm thick Ni film and Ni(100nm)/Nb(55nm) bilayer film. In neither case we observe the additional peak near the coercive field of Ni. 

\begin{figure}[hbtp!]
	\includegraphics[width=12 cm]{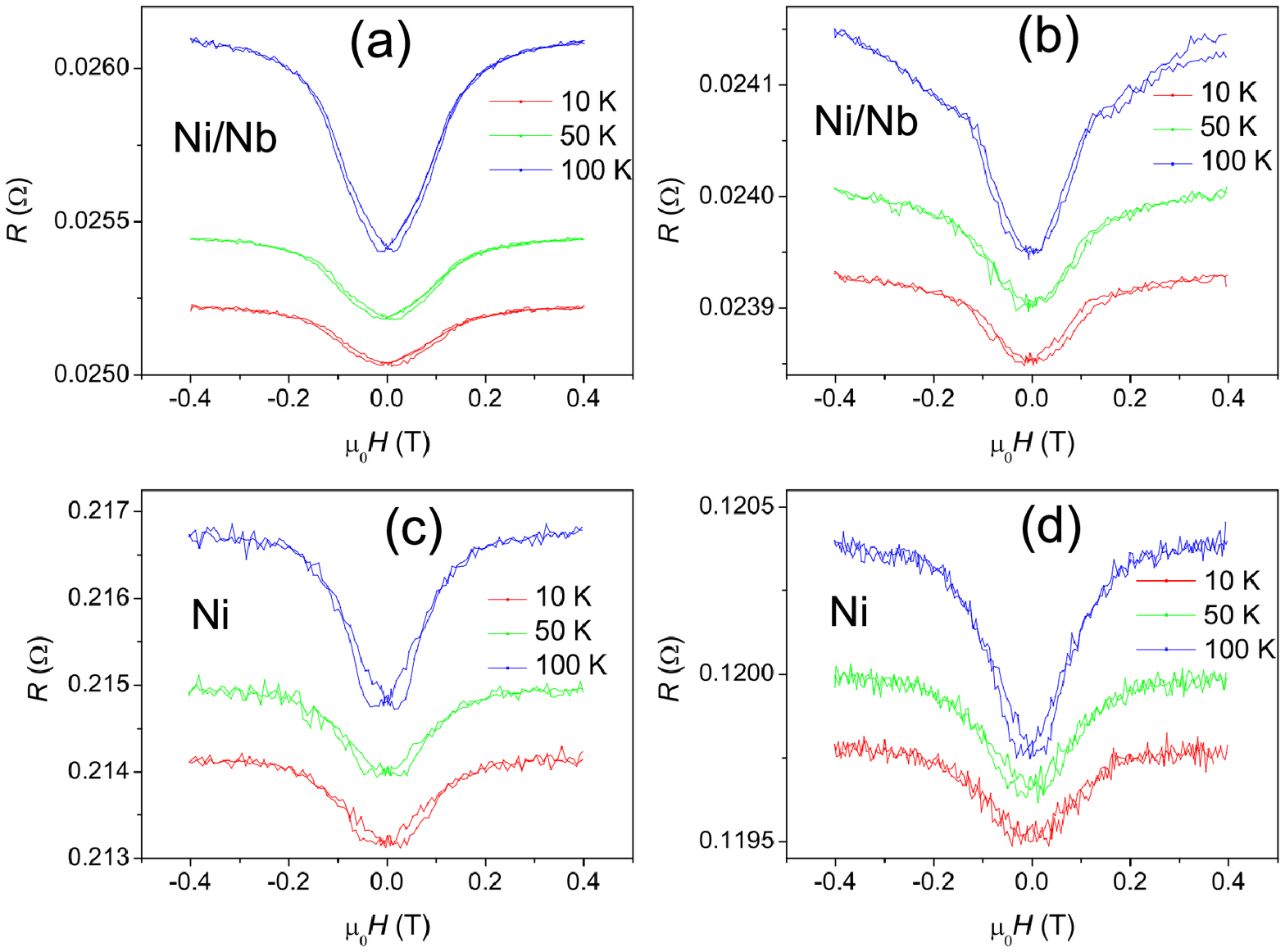}
	\caption{\textbf{Magneto-transport(MR) measurements of Ni(100 nm)/Nb(55 nm) bilayer and Ni(100 nm) thin film at various temperatures above the superconducting transition.}
		Panels (a) and (b) show the MR measurements at several temperatures for Ni/Nb bilayer in longitudinal (current along the magnetic filed) and transverse field orientations, respectively. Panels (c) and (d) show the same measurements of a single layer Ni film. All the curves are shifted along y axis for a clear comparison.	}
\end{figure}

\section{Estimation of domain wall magnetoresistance}

	Existing theories\cite{1,3} on DWMR allow to estimate the amplitude of MR peaks by using parameters from the magnetic hysteresis loop of the ferromagnetic layer. For the resistance due to a constant current passing through a single domain wall, in a constricted geometry, Ieda \textit{et al.}\cite{1} have found the following expression using the concept of spin accumulation:
 \begin{center}                      
	$R = 2P^{2}\rho_{F}\lambda_{F}A^{-1}F(\xi)$
\end{center}
Where P is polarization of conduction spin, $\rho_F $ is normal resistivity, $ \lambda_F $ is the spin-diffusion length, A is the cross sectional area of the constriction, and $ F(\xi) $ is a function of the ratio of domain wall width and spin-diffusion length. 
In our measurement geometry, current passes through many domain walls. Therefore, the amplitude of the DWMR peak is a sum of contributions from all the domain walls. Following Pati\~{n}o \textit{et al.}\cite{4}, we have used 	$n=n_{0}(1-|M(H)/M_{s}|)$ to find the number of domain walls, where $n_0$, $M$ and $M_s$ are the maximum number of domain walls at coercive field (H$_c$), the magnetization, and the saturation magnetization, respectively. 
\begin{figure}
\includegraphics[width=12 cm]{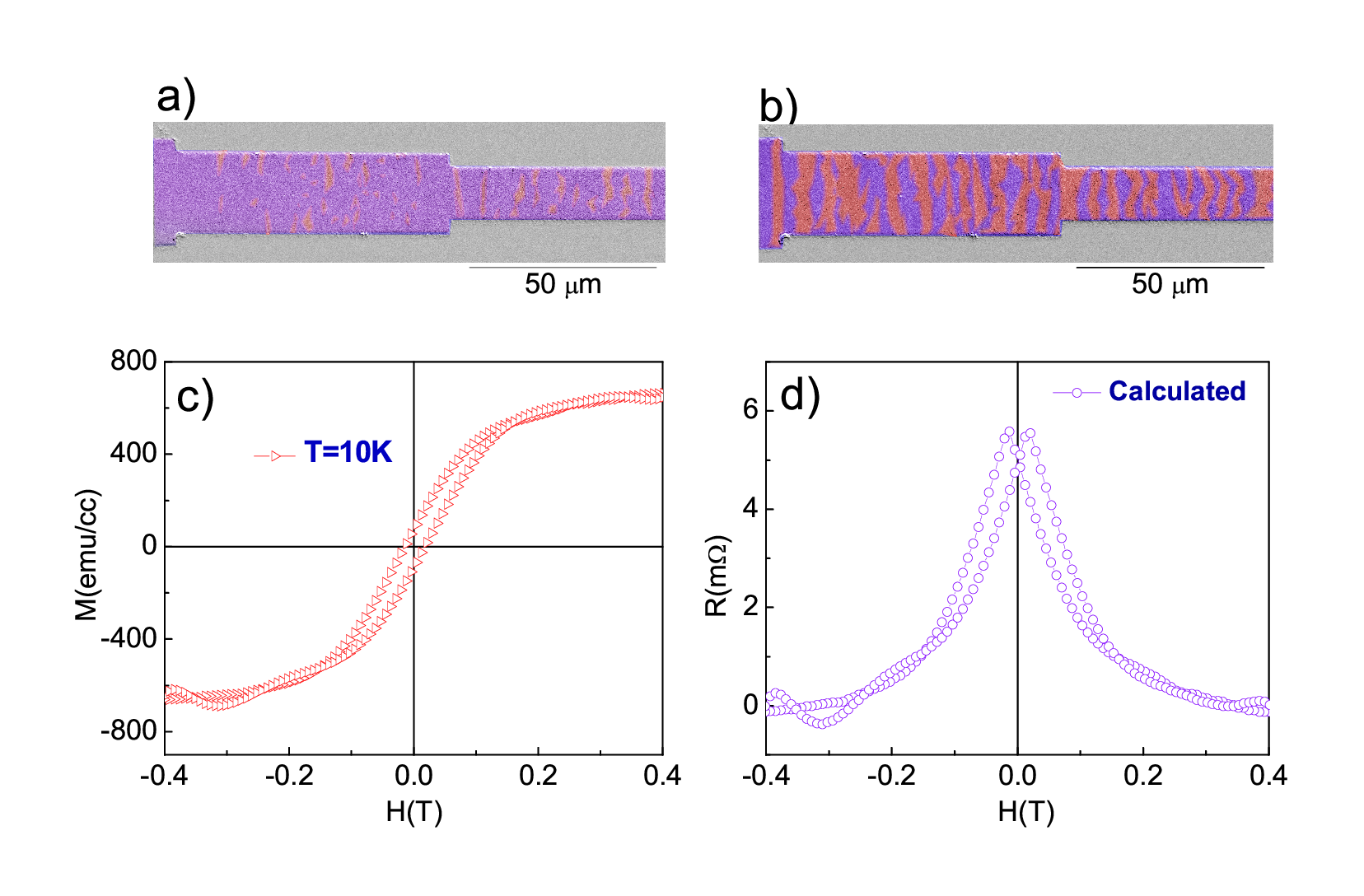}
	\caption{\textbf{Calculation of DWMR for perpendicular configuration from the magnetization versus in plane magnetic field curves for unpatterned Ni/Nb bilayers.} 
		fig.(a) and fig.(b) represent the kerr images showing increasing no. of domain walls on approaching the coercive field from the saturation field.
	    (c) panel c shows the MH measurements for Ni/Nb bilayers at a temperature of 10K for an in-plane applied magnetic field.
		(d) DWMR has been calculated by taking the magnetization values from MH measurement shown in panel c and by calculating the number of domain walls from the Kerr microscope images.}
\end{figure}
In Fig.S3(a) and S3(b) we compare the domain patterns in a nickel stripe near the saturation field and the coercive field, respectively. We notice that the number of domain walls is maximum near the coercive field. Fig. S3(c) shows the corresponding magnetization loop. Using $M$ and $M_s$ parameters from the hysteresis curve and $n_0$, the calculated DWMR is plotted in fig. S3(d). 
Therefore, the total resistance due to all the domain walls in Ni/Nb planar structure can be written as:
\begin{center}                      
	$R=R_{0}(1-|M(H)/M_{s}|)$
\end{center}

where                     $R_0 = n_{01}R_{01} + n_{02}R_{02}$

Here $R_{0}$ is the maximum resistance due to total number of domain walls present in the Ni/Nb planar structure at the coercive field, $n_{01}$ is the number of domain walls present in the minimum width stripe, $n_{02}$ is the number of domain walls present in the stripe adjacent to minimum width stripe, $R_{01}$ is the resistance caused by one domain wall present in the minimum width stripe when a constant current flows through it, $R_{02}$ is the resistance caused by one domain wall present in the adjacent stripe upon passing the same current through it. Here, $R_{01}$ and $R_{02}$ have been calculated using the Ieda's equation for DWMR. A spin polarization of P = 20$\%$, a spin diffusion length of $\lambda_{F}$ = 21nm, and normal resistivity value of, $\rho_{F}$ = 520 $n\Omega m$ has been used from previous experiements\cite{2,3}, for determination of $R_{01}$ and $R_{02}$. The value of parameter $ F(\xi) $ has been considered as 0.3 based on the previous reports\cite{2,3}. For the experimental geometry considered in this study, $n_{01}$ and $n_{02}$ have been calculated from the Kerr microscope images at coercive field for perpendicular configuration.

We have calculated DWMR using the above equation. The number of domain walls is maximum near the coercive field (H$_c$) and hence the DWMR is maximum at this field. This fact is evidenced in Kerr microscope images of patterned nickel stripes shown in Fig. S3(a) and Fig.S3(b), corresponding to magnetic field values much higher than H$_c$ and near H$_c$. An increase in number of domain walls was observed as one approaches the coercive field. This verifies the fact that the DWMR follows directly the number of domain walls. In fact, the peak amplitude of calculated DWMR corresponds to a value of 5.53 $m\Omega$ which is of the same order as the experimental value of 5.9 $m\Omega$. Thus, the calculated DWMR explains well the experimental magneto-transport measurements. There is a slight discrepancy in calculation and measurement in terms of magnetic field value corresponding to the MR peak. This can be explained from the fact that the magnetization values have been taken from the MH curve of a plain Ni/Nb bilayer. And, the value of coercive field increases slightly upon patterning the plane Ni/Nb bilayer. Therefore, the MR peaks appear at a lower magnetic field in calculation shown in Fig.S3(d) as compared to the measured one.

\section{Current dependence of domain wall magnetoresistance}

Here we plot the current dependence of domain wall magnetoresistance in the superconducting state (6K) of the Nb layer. With increasing current the Lorentz force on any nucleating vortices in the Nb layer would increase. Therefore, if a vortex locking-unlocking effect is the cause of the observed unconventional decrease in the DWMR then then the field at which the drop in DWMR happens would be different. However, this measurement (Fig. S4) shows that the drop in DWMR appears around the same magnetic field in all cases. 

\begin{figure}[htbp!]
	\includegraphics[width=11cm]{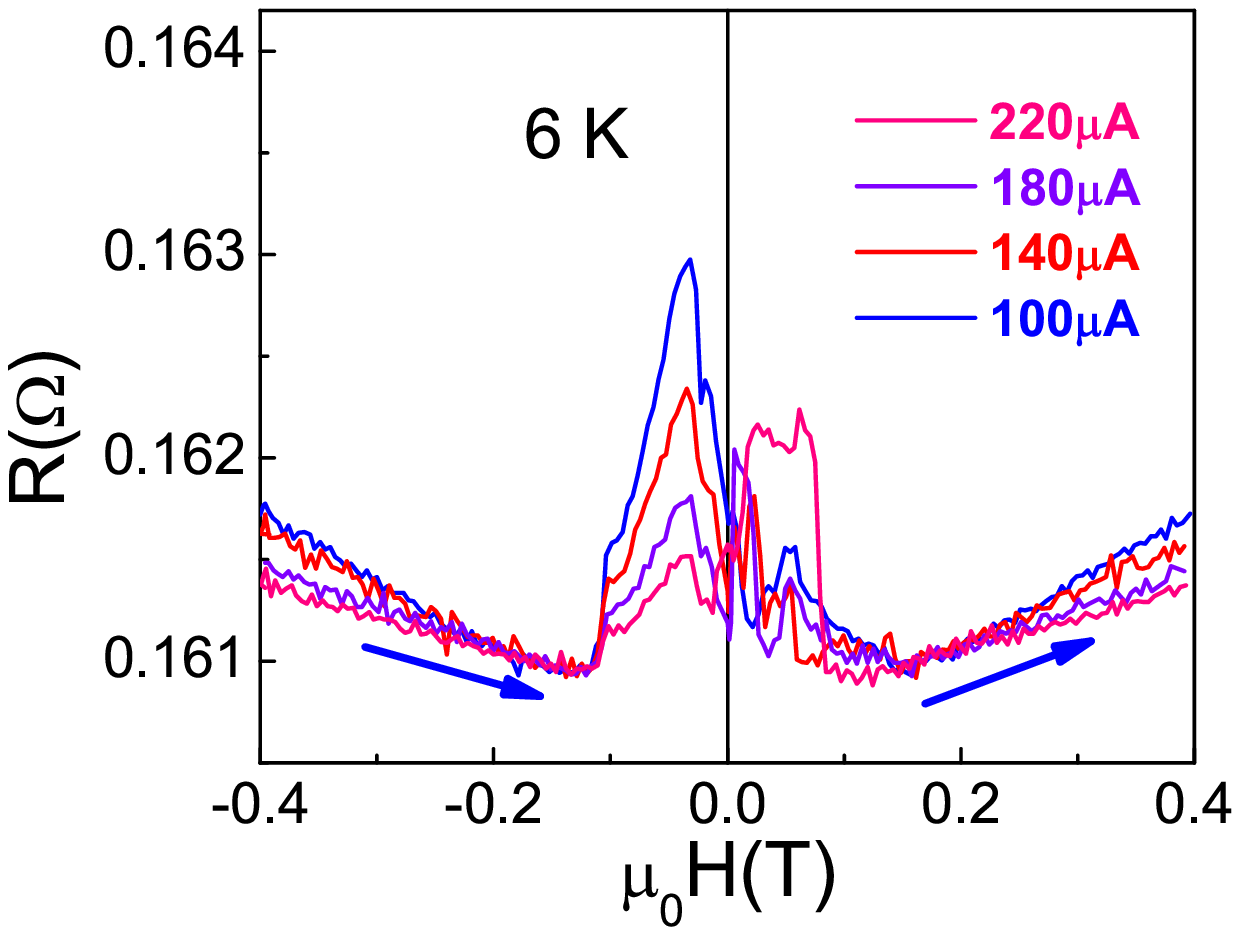}
	\caption{\textbf{Current dependence of DWMR below superconducting transition in transverse field configuration.} DWMR measurements performed at 6 K with several bias currents between 100 $\mu$m and 220 $\mu$m. decreases with increase in bias current at a temeprature of 6K. Arrows show the direction of magnetic field sweep.}
\end{figure}